\documentclass[aps,prd,showpacs,floatfix,nofootinbib,twocolumn,10pt]{revtex4}

\usepackage{color,amsmath,amssymb,graphicx,latexsym,subfigure}
\usepackage{threeparttable}

\begin{document}

\title{Testing the 130 GeV gamma-ray line with high energy resolution 
detectors}

\author{Ye Li$^{1,2}$ and Qiang Yuan$^3$}

\affiliation{
$^1$National Astronomical Observatories, Chinese Academy
of Sciences, Beijing 100012, P.R.China\\
$^2$National Astronomical Observatoires/Yunnan Observatory, Chinese 
Academy of Sciences, Kunming 650011, Yunnan, P.R.China\\
$^3$Key Laboratory of Particle Astrophysics, Institute of High Energy 
Physics, Chinese Academy of Science, Beijing 100049, P.R.China
}


\begin{abstract}

Recently some hints of the existence of $\gamma$-ray line around 130 GeV 
are reported according to the analysis of Fermi-LAT data. If confirmed it
would be the first direct evidence to show the existence of new physics
beyond the standard model. Here we suggest that using the forthcoming
high energy resolution $\gamma$-ray detectors, such as CALET and DAMPE,
we may test whether it is real line structure or just the background
effect. For DAMPE like detector with designed energy resolution 
$\sim1.5\%$, a line significance will reach $11\sigma$ for the same 
statistics as Fermi-LAT. For about $1.4$ yr survey observation, DAMPE
may detect a $5\sigma$ signal of such a $\gamma$-ray line.

\end{abstract}


\maketitle

\section{Introduction}

It was recently reported through an anlysis of the Fermi Large Area 
Telescope (Fermi-LAT) $\gamma$-ray data that there might be a monochromatic 
$\gamma$-ray line at energy $\sim 130$ GeV \cite{2012arXiv1203.1312B,
2012arXiv1204.2797W}. The significance is about $3-4\sigma$, taking into 
account the trial effect in the search. It was pointed out in Ref. 
\cite{2012arXiv1204.6047P} that the bump-like structure might coincide 
with the emission of the Fermi bubbles \cite{2010ApJ...717..825D,
2010ApJ...724.1044S}. In Ref. \cite{2012arXiv1205.1045T} the anthors 
re-analyzed the Fermi-LAT data and showed that the spatial distribution 
of the line emission should not correlate with the Fermi bubbles 
(see also Ref. \cite{2012arXiv1206.1616S}), but revealed some sub-structures.
The further analysis found the existence of abundant spectral 
features (excesses and dips) of the diffuse $\gamma$-rays and the
spectra suffered from spatial variations, which was against the dark
matter origin of the spectral structures \cite{2012arXiv1205.4700B}.

Several models employing dark matter annihilation/decay were proposed
to explain this tentative line structure \cite{2012arXiv1205.1520D,
2012arXiv1205.2688C,2012arXiv1205.3276C,2012arXiv1205.4151K,
2012arXiv1205.4675L,2012arXiv1205.4723R,2012arXiv1205.5789S}. See also
the expectation of the line emission \cite{2010JCAP...04..004J}. Although
the dark matter model may not easily to explain the above mentioned
features of the emission, especially the spatial distribution, it
would be meaningful to test such a line conjecture due to the potential
importance of a high energy $\gamma$-ray in the fundamental physics.
One way is to test the $\gamma$-ray emission from other targets where
dark matter is expected to be abundant, like the dwarf galaxies,
galaxy clusters and so on. However, the previous searches for line 
emission seem not sensitive enough to explore the recently reported
$130$ GeV line with annihilation cross section $\sim 10^{-27}$ cm$^3$
s$^{-1}$ (see e.g. \cite{2010PhRvL.104i1302A,2010JCAP...04..014A,
2011JCAP...05..027V,2012JCAP...04..030F}). The most recent search for
$\gamma$-ray line in the Milky Way can marginally exclude the dark
matter annihilation cross section $\sim 10^{-27}$ cm$^3$ at $130$ GeV
\cite{2012arXiv1205.2739F}. The constraint on dark matter annihilation
into $\gamma$-ray lines from observations of dwarf galaxies is relatively
weaker \cite{2012arXiv1206.0796G}. Another way is to use the accompanied
annihilation/decay to other final states to test the dark matter scenario
of the $\gamma$-ray line \cite{2012arXiv1205.6811B}. The latter one
depends on the detailed model of dark matter, however.

In this work we propose that this tentative $\gamma$-ray line structure
can be tested with the forthcoming high energy resolution $\gamma$-ray
detectors, such as CALorimeteric Electron Telescope (CALET\footnote{
http://calet.phys.lsu.edu/}) and DArk Matter Particle Explorer (DAMPE,
previously named TANSUO, \cite{Chang:DAMPE2011}). Both CALET and DAMPE
are dedicated to detect the cosmic ray nuclei, electrons and $\gamma$-ray 
photons in a wide energy band with very high energy resolution. Thanks 
to the very thick calorimeter of these detector ($30$ radiation lengths 
for CALET and $34.5$ radiation lengths for DAMPE), the energy 
resolution of them can reach or be better than $2\%$. As a comparison,
Fermi-LAT detector has only $10$ radiation lengths. The designed 
geometry factor of CALET is about $0.12$ m$^2$ sr, which is several 
times smaller than that of DAMPE ($\sim 0.5$ m$^2$ sr). In addition, the 
energy resolution of CALET is about $2\%$ at $\sim 100$ GeV, which is 
also a little bit worse than that of DAMPE ($\sim 1.5\%$). Therefore we 
will focus on the DAMPE mission in the following discussion. 

In the next section we will show the expected detectability of such a line 
emission on the DAMPE detector. Sec. III is the conclusion.

\section{Detectability of the line with high energy resolution detector}

\begin{figure*}[!htb]
\centering
\includegraphics[width=0.95\columnwidth]{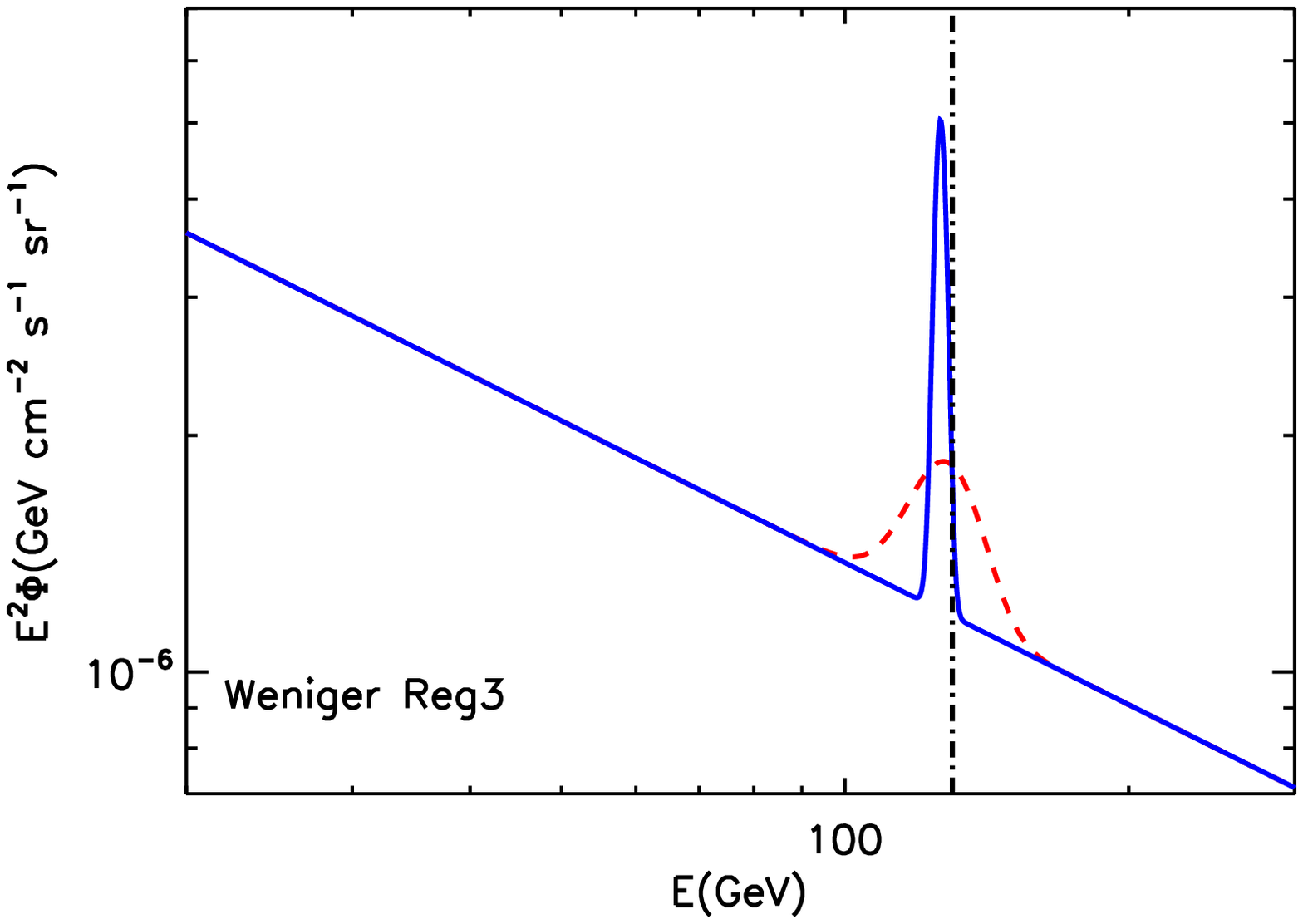}\\
\includegraphics[width=0.95\columnwidth]{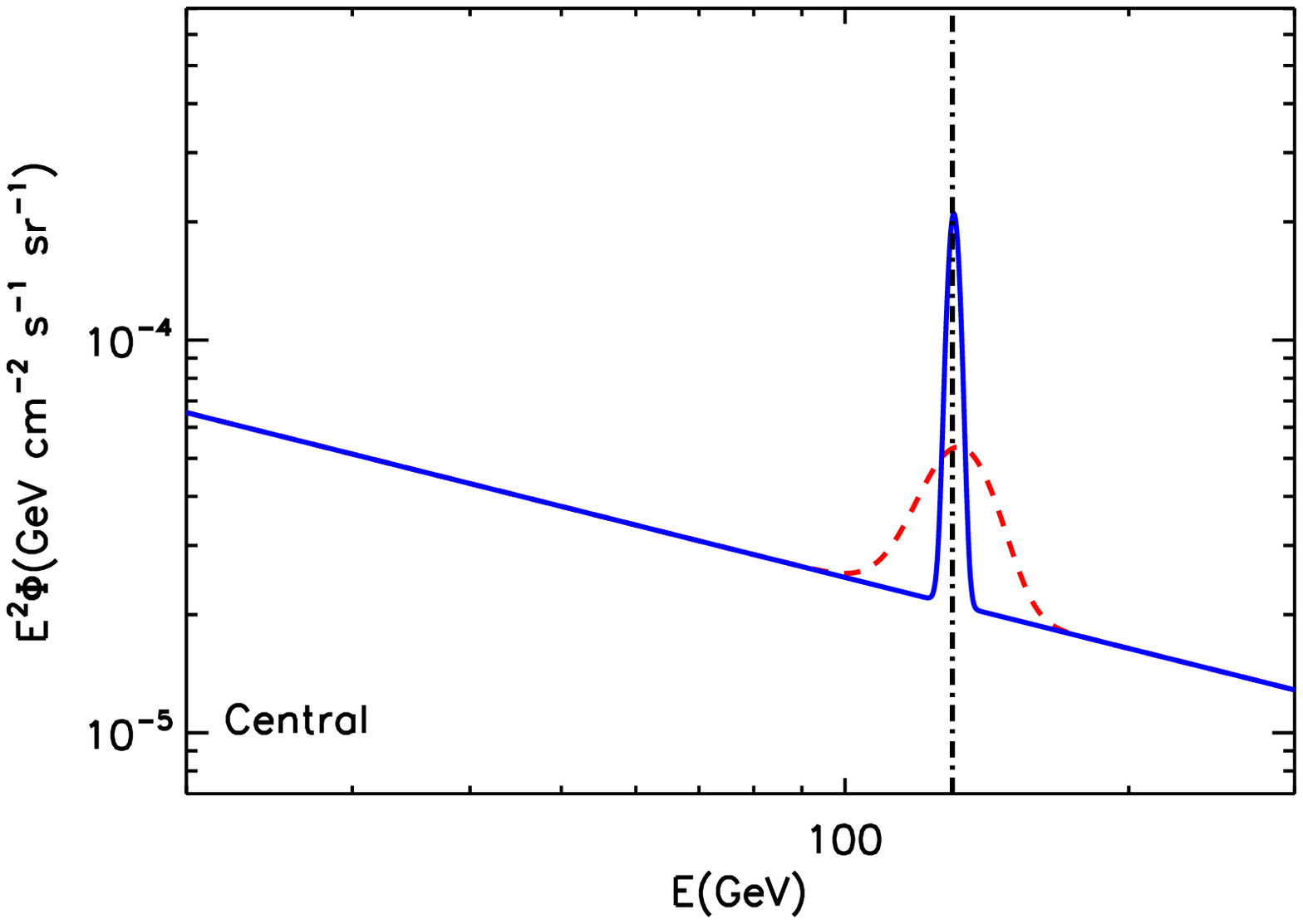}
\includegraphics[width=0.95\columnwidth]{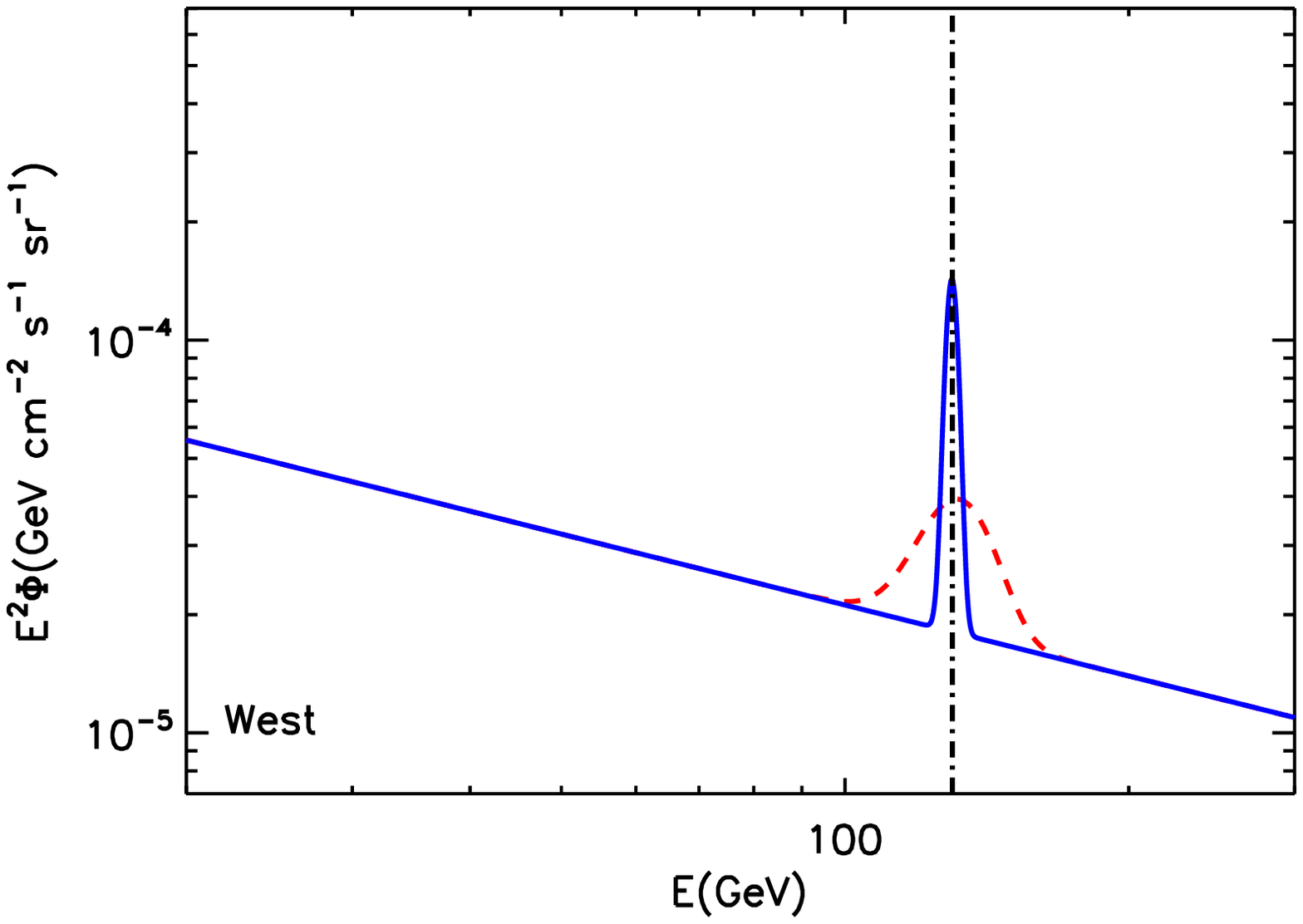}
\caption{The $\gamma$-ray spectra corresponding to Fermi-LAT energy 
resolution (dashed) and DAMPE energy resution (solid). The vertical
line in each panel shows the location of $130$ GeV.
\label{fig:spec}}
\end{figure*}

Here we choose several high significant regions shown in Refs.
\cite{2012arXiv1204.2797W,2012arXiv1205.1045T} for discussion, including
Reg3 of Weniger (2012), and the Central and West regions of Tempel
et al. (2012). The definition of the sky regions and energy spectra
of these three regions can be found in Refs. 
\cite{2012arXiv1204.2797W,2012arXiv1205.1045T}. The primary $\gamma$-ray
spectrum before reaching the detector is assumed to be power-law plusing
a monochromatic line. After convolving the energy spread function of 
Fermi-LAT detector, we fit the observed $\gamma$-ray spectra to determine
the primary spectra, including the energy and normalization of the 
$\gamma$-ray line. The energy spread function is assumed to be Gaussian, 
$$
f(E,E_0)=\frac{1}{\sqrt{2\pi}\sigma(E_0)}\exp\left[-\frac{(E-E_0)^2}
{2\sigma^2(E_0)}\right],
$$
where the Gaussian width $\sigma$ is the function of energy $E$.
A fit to the Monte Carlo simulation results presented in Fig. 4 of Ref. 
\cite{2012arXiv1205.2739F} gives $\sigma(E)/E\approx 0.0944-
0.0520\log_{10}(E)+0.0233[\log_{10}(E)]^2$. The fitted spectra 
corresponding to Fermi-LAT energy resolution are shown in Fig. 
\ref{fig:spec} (dashed lines).
 
For DAMPE detector the energy resolution is much higher, and we would
expect much sharper spectral feature if the $\gamma$-ray line is real.
The detailed simulation of the DAMPE detector is still on-going. 
In this work we employ a typical $1.5\%$ energy resolution for 
discussion. The expected energy spectra of the three sky regions
after convolving the DAMPE energy spread function are shown by the
solid lines in Fig. \ref{fig:spec}.

The energies of the $\gamma$-ray line for the three sky regions are fitted
individually since we have no prior information about it. The results are
slightly different from each other, and are shown in Table \ref{table1}.
We then calculate the number of photon events around the line energy
$E_0$. The photon number between $E_1$ and $E_2$ is
$$
N=\eta\,A_{\rm det}\,\Delta\Omega\,t_{\rm obs}\int_{E_1}^{E_2}\Phi(E){\rm d}E,
$$
where $\Phi(E)$ is the energy spectrum, $A_{\rm eff}$ is the effective area, 
$\Delta\Omega$ is the solid angle of the diffuse emission, $t$ is the 
observation time, and $\eta$ is the detection efficiency taking into 
account the physical efficiency, variation of effective area with photon 
arrival direction and the effective observational time of the survey mode. 
For Fermi-LAT observations, the parameter $\eta$ is determined by normalizing 
the resulting number to the detected one as listed in Table 1 of Ref. 
\cite{2012arXiv1205.1045T}. We integrate the spectrum within $\pm1\sigma$ 
range around $E_0$. For Fermi-LAT the width is about $0.089E_0$ and for 
DAMPE it is about $0.015E_0$. 

\begin{table*}[!htb]
\centering
\caption{Fitted line energy $E_0$, calculated photon event number within
$\pm1\sigma$ range around $E_0$, statistical significance and the expected
time to reach $5\sigma$ detection.}
\begin{tabular}{c|c|ccc|ccc}
\hline \hline
 Region &  $E_0$ & \multicolumn{3}{c}{Fermi-LAT} & \multicolumn{3}{|c}{DAMPE} \vspace{-0mm} \\
        &  GeV & $N_{\rm sig}/N_{\rm bkg}$ & significance & $t(5\sigma)$ & $N_{\rm sig}/N_{\rm bkg}$$^a$ & significance$^a$ & $t(5\sigma)^b$ \vspace{-0mm} \\
\hline
 Weniger Reg3 & $126.2$ & $24.2/52.5$ & $3.3$ & $8.6$ & $24.2/8.8$ & $8.2$ & $2.8$ \\
 Central      & $130.4$ & $17.1/13.5$ & $4.7$ & $4.2$ & $17.1/2.2$ & $11.4$ & $1.4$ \\
 West         & $129.8$ & $11.8/12.0$ & $3.4$ & $8.1$ & $11.8/2.0$ & $8.4$ & $2.6$ \\
\hline
\hline
\end{tabular}\\
$^a$Assuming the same exposure of Fermi-LAT and DAMPE;\\
$^b$The geometry factor of DAMPE is adopted to be half of that of Fermi-LAT.
\label{table1}
\end{table*}

The calculated numbers of events for the line component (signal) and power-law
component (background) are compiled in Table \ref{table1}. For DAMPE we
assume the same statistics as that of Fermi-LAT to show the numbers. It is
shown that with high energy resolution detectors, such a line emission
will be very significant ($>11\sigma$) and can be easily to be identified. 
We also give the expected time to reach $5\sigma$ detection for both
Fermi-LAT and DAMPE. Here all the parameters of DAMPE are assumed to be 
same as Fermi-LAT, except a factor of two smaller of the detection area
\cite{Chang:DAMPE2011}. For the Central region defined in Ref.
\cite{2012arXiv1205.1045T}, after about $4.2$ yr observation of 
Fermi-LAT the line will have a significance exceeding $5\sigma$. For
DAMPE the time will be only $1.4$ yr in spite that the effective area is 
smaller. Even if the line-like structure becomes more and more significant 
with the increase of the exposure of Fermi-LAT, it will still be difficult 
to distinguish from possible astrophysical origin of this spectral structure 
as discussed in Ref. \cite{2012arXiv1204.6047P}. However, for DAMPE 
experiment it will be much easier to identify whether it is a line or 
other spectral structure.

\begin{figure}[!htb]
\centering
\includegraphics[width=0.95\columnwidth]{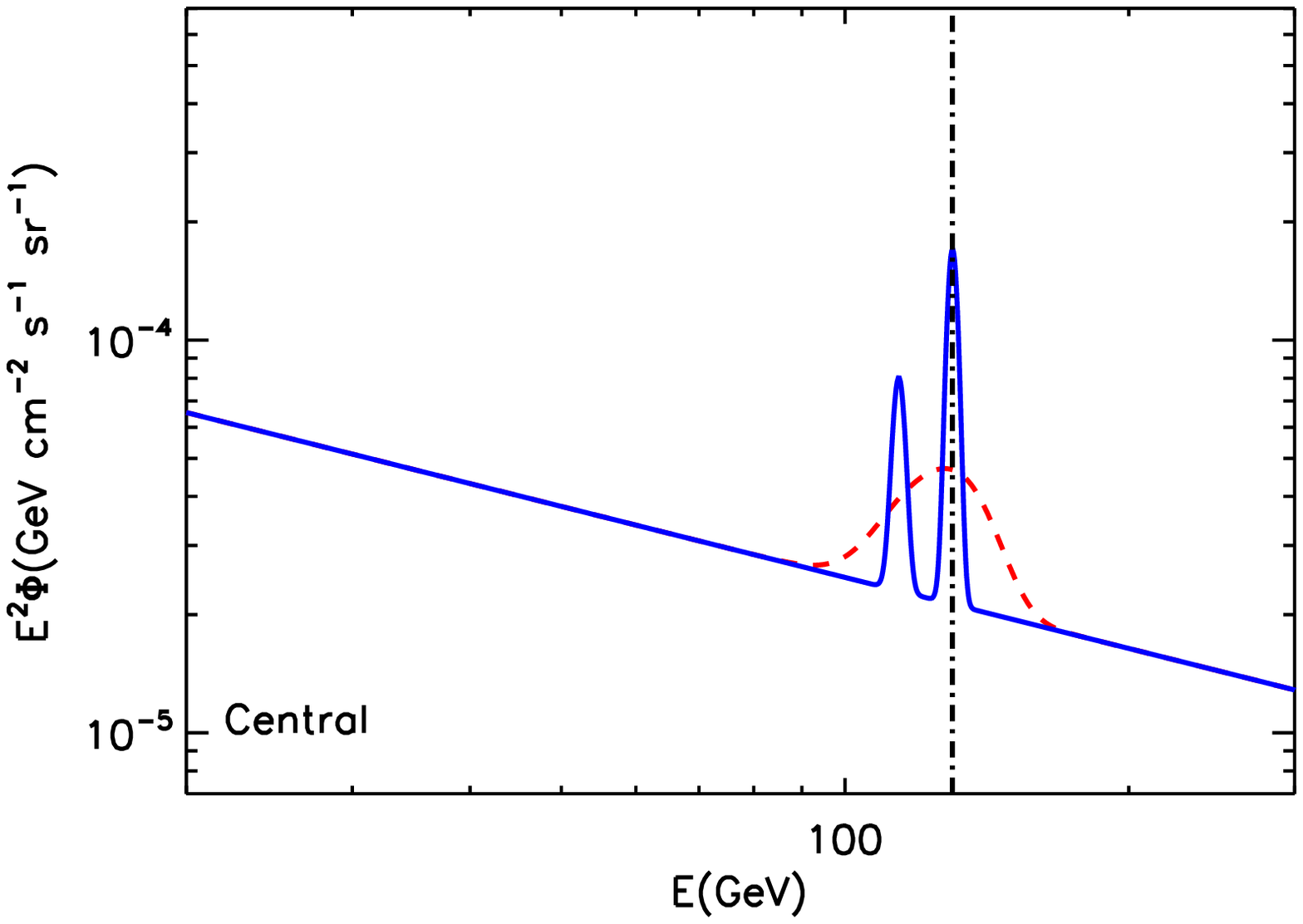}
\caption{The $\gamma$-ray spectra corresponding to two lines with
energies $130$ GeV and $114$ GeV, for Fermi-LAT energy resolution 
(dashed) and DAMPE energy resution (solid). The sky region is the
Central region of Ref. \cite{2012arXiv1205.1045T}.
\label{fig:2line}}
\end{figure}

It was also discussed that the current bump observed by Fermi-LAT might
consist with two lines and the detector can not resolve them due to
relatively bad energy resolution \cite{2012arXiv1205.4723R}. Given the
high energy resolution of DAMPE, we may have the potential to 
identify two lines if they contribute comparably to the flux.
We take the dark matter annihilation to $\gamma\gamma$ and $\gamma Z$
with equal branching ratios for example. For dark matter mass $m_{\chi}
=130$ GeV, the two $\gamma$-ray lines will have energies $130$ and $114$
GeV respectively, and the emissivity for $114$ GeV line will be two
times smaller than that of $130$ GeV line. The $\gamma$-ray spectra
for the Central region of Ref. \cite{2012arXiv1205.1045T} are shown
in Fig. \ref{fig:2line}. We can see that Fermi-LAT can not resolve
the two lines with $\sim 9\%$ energy resolution. The spectral profile
is a little broader than that of single line. At DAMPE the two
lines can be clearly seperated. Assuming the same statistics of the
current Fermi-LAT data, the two lines will have significances 
$8.3\sigma$ ($130$ GeV) and $3.7\sigma$ ($114$ GeV) respectively.

\section{Conclusion}

It is very interesting that some hints of $\gamma$-ray line around $130$ 
GeV were revealed by the Fermi-LAT data. It is not clear whether such
a spectral feature is really a line, or some astrophysical behavior, 
or just the fluctuation of background. In this work we propose to test
the line hypothesis with the forthcoming high energy resolution detectors.
It is shown that for a $\gamma$-ray detector with $1.5\%$ energy resolution,
the significance of this spectral structure will reach $\sim 11\sigma$
given the current statistics of photons, if it is indeed a line. Considering
the fact that the detector area of DAMPE is two times smaller than 
Fermi-LAT, we expect that after $\sim1.4$ yr operation of DAMPE in the
survey mode such a line structure will reach a $5\sigma$ detection
significance. For the fixed point observational mode it will be much
sooner to get a firm detection given the existence of such a $\gamma$-ray
line with high emissivity. We also discuss the potential to identify two 
nearby $\gamma$-ray lines as expected by some dark matter models by DAMPE.

\section*{Acknowledgements}

We thank Xiao-Jun Bi, Yi-Zhong Fan and Peng-Fei Yin for discussion.
This work is supported by National Natural Science Foundation of China 
under Grant No. 11105155.

\bibliographystyle{apsrev}
\bibliography{/home/yuanq/work/cygnus/tex/refs}

\end{document}